\documentclass[onecolumn,sort&compress,numbers]{els-mrw} 

\usepackage{amsmath,amssymb,amsfonts,amsthm,makeidx,graphicx}
\usepackage{txfonts}
\usepackage{helvet}

\usepackage[capitalize]{cleveref}

\begin{document}


\chapter{Meson Form Factors}
\label{chap:mesonformfactors}

\author[1]{Johan Bijnens}%


\address[1]{\orgname{Lund University}, \orgdiv{Division of Particle and Nuclear Physics, Department of Physics},
\orgaddress{Box 118, SE 221 00 Lund, Sweden}}

\articletag{Chapter}

\maketitle

\tableofcontents
\begin{abstract}[Abstract]
	We give an introduction to and a short overview of light meson form factors. We first discuss
	the classical picture and how it then fits in with amplitudes in quantum field theory. We give a short overview of the main theoretical methods and then discuss pion, kaon, eta and eta' form factors.
\end{abstract}

\begin{keywords}
 	meson\sep form factors\sep electromagnetic\sep weak
\end{keywords}

\begin{figure}[h]
	\centering
	\includegraphics[width=4cm]{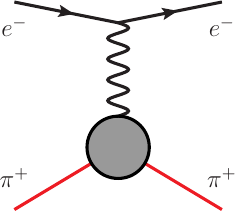}
	\caption{Electron-pion scattering with a pointlike photon-electron coupling
	 but the photon-pion coupling depends on its internal structure depicted by the shaded circle.}
	\label{fig:titlepage}
\end{figure}

\begin{glossary}[Nomenclature]
	\begin{tabular}{@{}lp{34pc}@{}}
		VMD & Vector Meson Dominance\\
		$\chi$PT & Chiral perturbation Theory
	\end{tabular}
\end{glossary}
\newcommand{\chpt}{$\chi$PT}

\section*{Objectives}
\begin{itemize}
	\item Form factors are a way to parametrize the effect of the strong interaction when mesons (or other strongly interacting particles) interact with external probes.
	\item Form factors have to be determined in order to extract fundamental parameters of the Standard Model.
	\item The main theoretical methods for the form factors are Chiral Perturbation Theory, dispersion relations and lattice QCD.
\end{itemize}

\section{Introduction}\label{intro}

The concept of a form factor goes back a long time and is used in molecular, atomic, nuclear and particle physics. Once one knows the scattering of an electromagnetic wave of a point charge one can determine the charge distribution by measuring the scattering as a function of momentum transfer $\vec Q$. The form factor is defined as the spatial Fourier transform of the charge distribution
\begin{align}
	\label{eq:atomic}
F\left(\vec Q\right)= \int d^3r \rho(\vec r) e^{i\vec Q\cdot\vec r}\,.
\end{align}
This is generalized to other interactions where one has some idea of the structure of the underlying amplitude following from Lorentz invariance and gauge or other symmetries. In particle physics they are most often used when one has a probe that does not have strong interactions which interacts with strongly interacting particles via a quark bilinear. The prime example is the photon that interacts with quarks via a term in the Lagrangian of the Standard Model as
\begin{align}
\label{eq:Aqbarq}	
\mathcal{L} = A^\mu(x) \sum_{q=u,d,s,c,b,t} e_q\bar q(x)\gamma_\mu q(x)\,.
\end{align}
But in addition to the vector current also form factors of other currents are considered and will be discussed below.
Due to the structure of \cref{eq:Aqbarq} the amplitude of electron-charged pion scattering to lowest order in the electromagnetic coupling as depicted in \cref{fig:titlepage} has the form
\begin{align}
	\label{eq:epitoepi}
A(e^-(q_1)\pi^+(p_1)\to e^-(q_2)\pi^+(p_2)) =
e^2 \bar e(q_2)\gamma^\mu e(q_1) \dfrac{1}{\left(q_2-q_1\right)^2}
\langle \pi^+(p_2)\vert j_\mu(0)\vert \pi^+(p_1)\rangle\text{ with }j_\mu(x) =  \sum_{q=u,d,s,c,b,t} e_q\bar q(x)\gamma_\mu q(x)\,.
\end{align}
The matrix element in \cref{eq:epitoepi} is to be evaluated in the strong interaction. Invariance under Lorentz transformations, rotations and translations strongly constrains the possible structure of the matrix element. The only quantities it can depend on are the Lorentz index $\mu$ as well as the four momenta $p_1$ and $p_2$. For particles with spin the polarization vector or spinor gives additional possibilities. We obtain
\begin{align}
	\label{eq:pivector}
	\langle \pi^+(p_2)\vert j_\mu(0)\vert \pi^+(p_1)\rangle = \left(p_2+p_1\right)_\mu F_V(q^2)+\left(p_2-p_1\right)_\mu F_{V2}(q^2)\,.
\end{align}
 The structure is dictated by the fact that the matrix element transforms under Lorentz transformations as a four vector with index $\mu$ and it can only depend on $p_2,p_1$ to produce that, that gives two possible structures, one with $p_{1\mu}$ and one with $p_{2\mu}$ that are multiplied by general functions of the Lorentz invariant quantities $q^2=\left(p_2-p_1\right)^2,p_1^2,p_2^2$. The latter two play no role since the charged pions are on-shell and thus $p_1^2=p_2^2=M_\pi^2$.

 We can now start adding additional general information on the form factors of \cref{eq:pivector}. First, the electromagnetic current $j_\mu$ is conserved, $\partial^\mu j_\mu=0$, and thus\footnote{In the remainder the point at which the current is applied will mostly be omitted.}
 \begin{align}
	\label{eq:pivectorconserved}
	\langle \pi^+(p_2)\vert \partial^\mu j_\mu(0)\vert \pi^+(p_1)\rangle
	= i(p_2-p_1)^\mu \langle \pi^+(p_2)\vert j_\mu(0)\vert \pi^+(p_1)\rangle
	= i\left[ \left(p_2^2-p_1^2\right) F_V(q^2) + \left(p_2-p_1\right)^2 F_{V2}(q^2) \right] =  i\left(p_2-p_1\right)^2 F_{V2}(q^2)\,. 
 \end{align}
This implies that $F_{V2}(q^2)=0$. The second general constraint follows from the fact that
\begin{align}
    \label{eq:pivectorquarks}
	\langle \pi^+(p_1)\vert \bar u\gamma_\mu u\vert \pi^+(p_1)\rangle = 2 p_{1\mu}\left(N_q-N_{\bar q}\right) = 2 p_{1\mu}\,,
\end{align}
i.e. the number of up-quarks minus the number of up-anti-quarks in the $\pi^+$ state and similarly for the down-quark\footnote{This is a general result from quantum field theory.}. The $\pi^+$ has one up-quark and one anti-down quark in addition to all possible quark-antiquark pairs and gluons which leads to the last identity in \cref{eq:pivectorquarks}. The consequence is that
\begin{align}
	\label{eq:pivector0}
F_V(q^2=0) = 1\,.
\end{align}

The charge radius\footnote{In the Breit frame with $p_1=(E,\vec Q/2), p_2 = (E,-\vec Q/2)$, we have that $q^2=\vec Q^2$ and the relation with \cref{eq:atomic} is most clear.} is defined via an expansion in $q^2$
\begin{align}
	\label{eq:radius}
F_V(q^2) = 1+\dfrac{r_V^2}{6} q^2+ c_V q^4+\cdots\,. 
\end{align}
The factor $1/6$ is conventional and follows from using \cref{eq:atomic} for a spherical charge contribution and expanding in $\vec Q$ giving $1-\langle r^2\rangle \vec Q^2/6+\ldots$ with $\langle r^2\rangle$ the average of $r^2$ for that charge distribution. Other parametrizations of the form factors that are often used are the monopole and dipole (not to be confused with the electric dipole form factor itself) versions
\begin{align}
	\label{eq:dipole}
F^\text{monopole}_V(q^2) &= \dfrac{1}{1-b q^2},&
F^\text{dipole}_V(q^2) &= \dfrac{1}{\left(1-d q^2\right)^2}\,.
\end{align}
The parameters $b$ and $d$ are often used to extract the radius as defined in \cref{eq:radius}.

It should be kept in mind that because of crossing symmetry the amplitudes for different processes are related. The amplitude \cref{eq:epitoepi} describes different processes. An incoming/outgoing electron can be replaced by an outgoing/incoming positron if the momenta are changed to minus themselves, the same goes for the positive charged and negatively charged pion. The kinematic ranges accessible are different for the different channels:
\begin{align}
\label{eq:tnegative}	
e^- \pi^+ \to e^- \pi^+&: -\infty \le q^2 \le 0\\
\label{eq:tpositive}
e^+ e^-\to \pi^+\pi^-&:  4 M_\pi^2 \le q^2 \le +\infty
\end{align}

Similar definitions are possible for many other amplitudes that are relevant and we will discuss a number of them below as well as refer to related encyclopedia articles. 

\section{Theoretical methods}
\label{sec:theory}

The form factors as defined in \cref{eq:pivector} and their generalizations give us information about the structure of the states involved. This can be obtained from experiment and from theory and the comparison of the two is a test of the theory of the strong interaction. The number of form factors and their dependence on kinematic quantities is determined by the Lorentz structure of the amplitude in question. The meson form factors have also been studied in many different theoretical approaches.

\subsection{Chiral perturbation theory}

The QCD Lagrangian has a chiral symmetry which is believed to be spontaneously broken \cite{Nefediev:2025zkv}. The effective field theory based on this is called Chiral Perturbation Theory and is one of the main underlying theoretical metods for light meson form factors. It is described in more detail in \cite{Meissner:2024ona}.

\subsection{Dispersive}

Form factors also have to obey dispersion relations. This combined with Watson's theorem on the phase of the form factors allows to obtain strong dispersive results on many form factors. Relevant articles in the encyclopedia are the coupled channel article \cite{Oller:2025leg}, the dispersive methods article \cite{Colangelo:2025sah} and the dispersion relations article \cite{Kubis:2025zji}.

\subsection{Lattice QCD}

Many quantities in hadronic physics are now calculated directly from lattice QCD, this includes the form factors discussed in this article. This is described in the articles related to lattice QCD.

\subsection{Modelling}

The quantities discussed here are treated also in models of low-energy QCD. This includes constituent quark models, the NJL model and extensions and various functional approaches\footnote{The Dyson-Schwinger equations become models when approximations are included.} \cite{Eichmann:2016yit}. Resonance saturation in the form of Vector meson Dominance \cite{Sakurai:1960ju} and its modern descendant resonance chiral theory \cite{Ecker:1988te,Ecker:1989yg} fall also in this category. We do not discuss these approaches.

\subsection{Perturbative QCD}

Form factors can be treated directly from QCD also with analytical methods. There are two main parts here. The first one is the use of QCD sum rules and especially light cone sum rules \cite{Colangelo:2000dp} and the second the use of Efimov-Radyushkin-Brodsky-Lepage studies of the form factors \cite{Lepage:1980fj,Efremov:1978rn}. These methods have not been included in the present overview article.

\section{Pion form factors}

\subsection{Vector form factor}

The pion vector form factor $F_V(q^2)$ defined in \cref{eq:pivector} is the oldest and by far most studied meson form factor. A very old way to study this is the use of Vector meson Dominance (VMD) \cite{Sakurai:1960ju} where the coupling to the photon is mediated by an intermediate vector meson. This was further developed in \cite{Gounaris:1968mw} in a parametrization which is still heavily used today, the Gounaris-Sakurai parametrization. At low energies the form factor was calculated in \chpt{} in \cite{Gasser:1983yg} in two-flavour \chpt{} and in \cite{Gasser:1984ux} in three-flavour \chpt{}. This was extended to two-loops in \cite{Bijnens:1998fm} and \cite{Bijnens:2002hp}. More recently the main approach has been to include dispersion theory to get also at the pion form factor at higher $q^2$. A recent paper where further references can be found is \cite{Colangelo:2018mtw}. 

On the experimental side there has been a lot work done recently, especially regarding the contribution to the muon anomalous magnetic moment. Unfortunately the positive $q^2$ part here suffers from experiments that are incompatible within their quoted errors, see the discussion in \cite{Aliberti:2025beg}. On the spacelike side, negative $q^2$, the measurement are still dominated by the NA7 experiment \cite{NA7:1986vav}.

A similar form factor is needed for pion beta decay ($\pi^+\to\pi^0 e^+\nu$) which provides an alternative method to determine $V_{ud}$. Here corrections to the isospin relation from $F_V(q^2)$ are the most important theoretical restriction as discussed in \cite{Cirigliano:2002ng}. Radiative corrections have been studied more recently in \cite{Seng:2020jtz}.

The charge radius gives an indication of the size of the pion \cite{ParticleDataGroup:2024cfk}
\begin{align}
	\label{eq:pichargeradius}
r_V^\pi = 0.659 \pm 0.004~\textrm{fm}
\end{align}

\subsection{Scalar form factor}

The scalar form factor is defined by
\begin{align}
\langle \pi^+(p_2) \vert \bar u u+\bar d d\vert\pi^+(p_1)\rangle = F_S(q^2)\,.
\end{align}
It cannot be measured directly but it can be determined from dispersion theory or lattice QCD. The most recent determination from lattice QCD is in \cite{vonHippel:2025uhr} and from dispersive in \cite{Ropertz:2018stk}. Earlier work can be traced from here. The \chpt{} calculations are done at one-loop in \cite{Gasser:1983yg,Gasser:1984ux} and at two-loop in \cite{Bijnens:1998fm,Bijnens:2003xg}. A major breakthrough for the scalar form factor was the study of \cite{Ananthanarayan:2004xy}. The scalar form factor behaves quite differently from the vector form factor. It is not dominated by a single resonance as the electromagnetic form factor and the radius is also different. 

The scalar radius is larger than the vector one \cite{Ananthanarayan:2004xy}
\begin{align}
r_S^\pi = 0.78 \pm 0.02~\textrm{fm}\,.
\end{align}
One conclusion here is that the size of composite particles depends on how it is defined.

\subsection{Transition form factor}
\label{sec:piTFF}

The pion transition form factor is defined as
\begin{align}
	\int d^4 x e^{i q_1\cdot x}\langle 0\vert T j_\mu(x) j_\nu(0)\vert\pi^0(q_1+q_2)\rangle
	= -\epsilon_{\mu\nu\alpha\beta} q_1^\alpha q_2^\beta F_{\pi^0\gamma^*\gamma^*}(q_1^2 ,q_2^2)\,.
\end{align}
The one-loop calculation in \chpt{} was done in \cite{Bijnens:1988kx}. A full dispersive treatment was done in \cite{Hoferichter:2014vra}. A major use of this transition form factor is the determination of the pion pole contribution to the muon anomalous magnetic moment \cite{Hoferichter:2018kwz}. 
The most recent measurement is \cite{BESIII:2025zjx} and ditto lattice QCD calculation \cite{ExtendedTwistedMass:2023hin}.

\subsection{$\pi_{\ell2\gamma}$}

The decay $\pi\to\ell\nu\gamma$ is described by two form factors and a part that is determined by $F_\pi$, the pion decay constant. The form factor part is done by looking at the amplitudes
\begin{align}
\int d^4 x e^{i q_1\cdot x}\langle 0\vert T j_\mu(x)\bar d \gamma_\nu (1-\gamma_5)u(0)\vert\pi^+(q_1+q_2)
	= \epsilon_{\mu\nu\alpha\beta} q_1^\alpha q_2^\beta F_V(q_1^2 ,q_2^2)
	  + F_A(q_1^2,q_2^2)(q_1\cdot q_2 g_{\mu\nu}-q_{1\mu }q_{2\nu})
	\,.
\end{align}
The decay is evaluated at $q_1^2=0$. $F_V(0,q_2^2)$ is related by isospin to the vector transition form factor.
The one-loop result in two flavour \chpt{} for $F_A$ is in \cite{Gasser:1983yg} and at two-loops in \cite{Bijnens:1996wm}, the one-loop result for $F_V$ is in \cite{Bijnens:1996wm}. The three flavour result for $F_A $ at one-loop is in \cite{Gasser:1984ux} and at two-loop in \cite{Geng:2003mt}. 
Radiative corrections can be found in \cite{Unterdorfer:2008zz}. Lattice QCD has so far only used the isospin relation w.r.t. the transition form factor.
\section{Kaon form factors}

\subsection{Vector form factors}

\subsubsection{Electromagnetic form factor}

The kaon electromagnetic form factor is defined analogous to \cref{eq:pivector} but with the charged and neutral kaon rather than the $\pi^+$ in the matrix element. The \chpt{} calculation at one-loop order is in \cite{Gasser:1984ux} and at two-loop order in \cite{Bijnens:2002hp}. The dispersive study has been updated recently in \cite{Stamen:2022uqh}.

The experimental measurement of the kaon radius \cite{ParticleDataGroup:2024cfk} is
\begin{align}
r_V^{K^+} = 0.560 \pm 0.018~\textrm{fm}
\end{align}
in reasonable agreement with the more accurate dispersive result \cite{Stamen:2022uqh}
\begin{align}
r_V^{K^+} = 0.599 \pm 0.002~\textrm{fm}\,.
\end{align}

\subsubsection{$K_{\ell 3}$}

The decays $K\to\pi\ell\nu$ are described by two form factors for each possible decay
\begin{align}
\langle\pi(p_2)\vert \bar s \gamma_\mu u\vert K(p_1)\rangle = (p_2+p_1)_\mu f_+(q^2)+(p_1-p_2)_\mu f_-(q^2)
\end{align}
with $q^2=(p_2-p_2)^2$. 
The axial part of the weak current does not contribute to this decay since the strong and electromagnetic interactions conserve parity. 
Alternatively one uses the parametrization in in terms of $f_+$ and the scalar form factor $f_0$
\begin{align}
f_0(q^2)\equiv f_+(q^2)+\dfrac{q^2}{M_K^2-M_\pi^2}f_-(q^2).	
\end{align}
The reason is that using $i\partial^\mu  ( \bar s \gamma_\mu u) = (m_u-m_s) \bar s u$, we can see that $f_0$ is proportional to the scalar form factor for the same transition. $f_+(0)$ is important in the determination of $V_{us}$. The one-loop calculation including isospin breaking from quark masses is in \cite{Gasser:1984ux,Leutwyler:1984je}. The two-loop result is worked out in \cite{Bijnens:2003uy} and including the up-down mass difference in \cite{Bijnens:2007xa}.
Electromagnetic corrections are important and are studied in detail in \cite{Cirigliano:2008wn}. 
This was put together in \cite{Kastner:2008ch}. The radiative corrections have been updated in \cite{Seng:2022wcw}.

These form factors have been calculated by many lattice QCD collaborations. An overview and discussion can be found in \cite{FlavourLatticeAveragingGroupFLAG:2024oxs}.

\subsection{Scalar form factors}

The scalar form factors of kaons are defined as
\begin{align}
\langle K(p_2) \vert \bar q q(0) \vert K(p_1)\rangle = F_S(q^2)
\end{align} 
with again $q^2=(p_2-p_1)^2$. These exist for $q=u,d,s$ and $K=K^+,K^0$. They are known at one loop in \chpt{} since \cite{Gasser:1984ux} and at two-loop from \cite{Bijnens:2003xg}. They are not measured directly experimentally but can be determined from dispersion relations for $\pi K$ scattering and then doing a coupled channel analysis \cite{Buettiker:2003pp,Pelaez:2018qny,Danilkin:2020pak,Cao:2025hqm}.

\subsection{$K_{\ell4}$}

The decay $K\to\pi\pi\ell\nu$ has four form factors, $F,G,R$ and $H$, as defined by
\begin{align}
	I_\mu &= \frac{1}{\sqrt2}\langle \pi^i(p_1)\pi^j(p_2)\vert \bar s \gamma_\mu(1-\gamma_5)u\vert K^k(p)\rangle\\
   I_\mu &= -\dfrac{i}{M_K}\left[F\left(p_1+p_2\right)_\mu+G\left(p_1-p_2\right)_\mu+R\left(p_\ell+p_\nu\right)_\mu\right]
   -\dfrac{H}{M_K^3}\epsilon_{\mu\nu\rho\sigma}(p_\ell+p_\nu)^\nu(p_1+p_2)^\rho(p_1-p_2)^\sigma
\end{align}
This decay happens for $ijk = +-+,00+,0-0$.

The form factor $H$ is anomalous and is known in \chpt{} to one-loop \cite{Ametller:1993hg}. The others have been calculated at one-loop in \cite{Bijnens:1989mr,Riggenbach:1990zp,Bijnens:1994ie} and to two-loops in \cite{Amoros:2000mc}.
Isospin breaking effects, including QED, can be traced from \cite{Stoffer:2013sfa}. The most recent full dispersive analysis is in \cite{Colangelo:2015kha}. Historically this decay has played a major role in determining the $\pi\pi$ scattering lengths since it is one of the few good sources of low-energy pion pairs. The best measurements have been performed by NA48 \cite{NA482:2007xvj}.

\subsection{$K_{\ell2\gamma}$}

This is the kaon semileptonic decay equivalent of the the $\pi_{\ell2\gamma}$ decay. In exactly the same way it has two form factors, a vector and an axial one. The vector decay one is known to one-loop in \chpt{} \cite{Ametller:1993hg} and the axial one from \cite{Gasser:1984ux}. The axial one is known to two-loop order from \cite{Geng:2003mt}. This decay is one of the places where the sign of the anomaly can be checked due to the interference between the two form factors in the decay rate. This decay has also started being calculated in lattice QCD, see e.g. \cite{DiPalma:2025iud} and references therein.

\subsection{$K_{\ell3\gamma}$}

This is the decay $K\to\pi\ell\nu\gamma$ and there are 7 form factors in the general case, 4 for the vector current and 3 for the axial current decay. the \chpt{} calculation to order $p^4$ is done in \cite{Bijnens:1992en}. There is in general good agreement between the measurements and the \chpt{} calculations.

\section{Eta/Eta' form factors}

\subsection{Transition form factors}

These form factors are defined as in \cref{sec:piTFF} but with the $\eta$ and $\eta^\prime$ rather than the $\pi^0$ in the amplitude. The one-loop \chpt{} calculation was done long ago in \cite{Bijnens:1988kx}. Dispersive estimates have been done by many people. The most recent is \cite{Holz:2024diw}. A discussion of the lattice QCD calculations can be found in \cite{Aliberti:2025beg}.

\subsection{Vector form factor}

Similar to the kaon decay the eta vector form factor plays a role in the decay $\eta\to\pi\ell\nu$. The decay violates isospin and thus isospin violation from the quark mass difference $m_u-m_d$ and QED needs to be taken into account. This was done in \chpt{} in \cite{Neufeld:1994eg}.
An update for the crossed amplitude in tau decay is \cite{Garces:2017jpz}.

\section{Others}

The main part of this overview has been the light meson form factors. The same methods can be used for all amplitudes with external currents. The baryon case is discussed in \cite{Lin:2024rak} but also vector meson form factors can be defined. An example is the omega transition form factor. Here the dispersive study is in \cite{Schneider:2012ez}.



\section{Conclusions}
\label{sec:conclusions}
The light meson form factors discussed here are interesting observables by themselves allowing to study a large number of properties of the strong interaction. In addition they play a major role in measuring Standard Model parameters to high precision. One possible use is to determine the sizes of bound states where one needs to keep in mind that the sizes or radii depend on the probe or current used.

The form factors by themselves also give tests of QCD when they are calculated as well as measured.


\begin{ack}[Acknowledgments]%
I thank the many collaborators I have had over the years for calculating especially the \chpt{} form factors. A non exhaustive list is G. Amoros, C. Bernard, A. Bramon, G. Colangelo, F. Cornet, P. Dhonte, E. G\'amiz, K. Ghorbani, K. Kampf, J. Relefors and  P. Talavera. The number of people I have discussed with and whose work has been useful in understanding form factors is simply too large to list them.
\end{ack}


\bibliographystyle{Numbered-Style} 
\bibliography{reference}

\end{document}